\documentclass[review]{elsarticle}

\usepackage{hyperref}

\journal{Journal of \LaTeX\ Templates}

\usepackage{numcompress}
\begin{document}

\begin{frontmatter}

\title{Mean and Gaussian Curvature of Lipid Mesophases measured using molecular dynamics}

\author[Physics]{Christopher Brasnett}

\author[Physics,BCFN]{Annela Seddon\corref{CA}}
\cortext[CA]{Corresponding author}
\ead{annela.seddon@bristol.ac.uk}

\address[Physics]{School of Physics, University of Bristol, Tyndall Avenue, BS8 1FD}
\address[BCFN]{Bristol Centre for Functional Nanomaterials, School of Physics, University of Bristol, Tyndall Avenue, BS8 1FD}

\begin{abstract}
The ability to measure the mean and Gaussian curvature in lipid mesophases is important in our understanding of their formation and properties, and can be achieved both experimentally and computationally. Here we present a method to measure curvature for any mesophase using Molecular Dynamics simulations. However the reliability of the results is highly dependent on the choice of fitting algorithm, the atoms within the lipid membrane that are selected for fitting and the number of atoms that are included in the fit. We compare the results of our method to the L${\alpha}$ mesophase as previously studied, and subsequently extend the method to the Q$_{II}^{D}$ and H$_{II}$ mesophase, not previously studied in this way computationally, but whose curvatures are known both analytically and experimentally. By systematically comparing our results, we demonstrate a robust method which can be used in general to measure curvature.
\end{abstract}


\end{frontmatter}


\section{Introduction}

Membrane curvature is a fundamental property of membrane systems, and is known to play a role in membrane fusion, membrane protein activity, and lateral membrane organisation  \cite{Kawamoto2015, ChangIleto2011, Bigay2005,Lee2005, Lundmark2008,vanMeer2008,CallanJones2013,Baoukina2018}. The curvature of membranes arises from the global restraint of molecular composition, and is a key factor when considering the elasticity of membranes\cite{hyde1997the}. The mean (H) and Gaussian (K) curvatures of a surface are defined by the mean and product of the principal curvatures respectively. The Helfrich Hamiltonian, describing the energy density of membranes, is dependent on both measures, indicating that the dynamic behaviour - of both model and real membrane systems - is significantly influenced by the underlying geometry of the surface around which the lipid bilayer is formed\cite{Helfrich1973}.

Lipid mesophases may be characterised by their mean and Gaussian curvatures, from the flat planar lamellar (L$_{\alpha}$, Figure \ref{fig:molecules}d)), to the highly curved inverse hexagonal (H$_{II}$, Figure \ref{fig:molecules}f)), where cylindrical water channels are arranged in a hexagonal pattern. One of the most studied non-planar mesophases is the Diamond cubic phase (Q$_{II}^{D}$, Figure \ref{fig:molecules}e)), based on triply periodic minimal surfaces, mathematical surfaces defined by zero mean curvature, and negative Gaussian curvature \cite{PhysRevE.59.5528}. L$_{\alpha}$ mesophases have been extensively studied \textit{in silico} for their physiological importance \cite{Risselada2008,Carpenter2018,Bennett2013}. 

Experimentally in sufficiently large and simple systems, curvatures can be calculated directly, for example using microscopy or dynamic light scattering\cite{VillamilGiraldo2020, Larsen2017}. For a spherical liposome, the principal curvatures are identical, and the curvature is simply the reciprocal of the radius experimentally measured. However, for mesophases with more complex topologies, such as the inverse bicontinuous cubic phases, curvatures are dependent on the dimensions of the unit cell\cite{Tang2014}. 

Non-lamellar mesophases, on the other hand, have been the subject of fewer computational studies, especially in comparison to experimental studies where they have been studied extensively, particularly using small angle X-ray scattering techniques to investigate the nature of mesophase transitions\cite{Seddon2006, Seddon2014, Squires2000, vantHag2017, Conn2006}. However, in recent years, there has been increased interest in the simulation of non-lamellar lipid mesophases\cite{Khelashvili2012, Johner20140, Johner20141}. This has been  driven by the need to understand their role in processes such as drug delivery, templating, and membrane protein crystallisation \cite{Barriga2019, Zhai2019, Richardson2017, Zabara2018, Caffrey2015}. One less-studied computational aspect is the role of curvature in these processes. \citeauthor{Yesylevskyy2014} developed a method to measure curvature from simulated mesophases based on methods from computer graphics, investigating its use on a planar lipid system containing varying quantities of cholesterol \cite{Yesylevskyy2014}. However, this method has not yet been applied to non-lamellar systems. As mentioned above, membrane curvature plays a significant role in membrane protein activity and function, with membrane proteins themselves having complex mechanisms with which to sense the curvature of the membrane in which they reside \cite{Tonnesen2014, Lundbk2009, McMahon2015, Jarsch2016, Gov2018,Prinz2009,Zimmerberg2005}. Moreover, the lipid composition of membranes themselves can both induce, and be sorted by curvature \cite{Black2014,Cheney2017,CallanJones2013, Sorre2009}. Understanding the method of \citeauthor{Yesylevskyy2014} in mesophases with non-trivial curvatures as we present here will therefore allow us to further our understanding of the effects that curvature has on intra-membrane organisation.

To examine these phenomena, here, we explore the method by evaluating its performance on three lipid mesophases. We firstly investigate how the method corroborates using different fitting methods, to ensure a high level of reproducibility. Further to that, we assess for the first time how the use and implementation of the method varies in different mesophases, measuring curvature with different system parameters. Of the mesophases used in this study, two have non-zero curvatures, the H$_{II}$ and the Q$_{II}^{D}$. The Q$_{II}^{D}$ and the H$_{II}$ are formed spontaneously in water by monoolein (MO) and dioleoylphosphoethanolamine (DOPE) respectively. The third mesophase is the L$_{\alpha}$ mesophase, with both mean and Gaussian curvatures equal to zero, formed by dipalmitoylphosphatidylcholine (DPPC) (Figure \ref{fig:molecules}a)). Coarse grained figures of the three molecules and the mesophases they form in water are shown in Figure \ref{fig:molecules}. 

\section{Methods}

\subsection{Simulation details}

\begin{table}[ht]
\begin{tabular}{c|cccc}
\begin{tabular}[c]{@{}c@{}}Mesophase \\ and lipid\end{tabular} & \begin{tabular}[c]{@{}c@{}}Number\\ of water molecules\end{tabular} & \begin{tabular}[c]{@{}c@{}}Number \\ of lipids\end{tabular} & \begin{tabular}[c]{@{}c@{}}Weight proportion\\ of water\end{tabular} & \begin{tabular}[c]{@{}c@{}}Lattice parameter\\ (Å)\end{tabular}               \\ \hline
H$_{II}$ (DOPE)                                                & 1328                                                       & 300                                                         & 0.3                                                                  & 50.3 $\pm$ 0.3                                                                    \\
L$_{\alpha}$ (DPPC)                                            & 768                                                        & 128                                                         & 0.37                                                                 & \begin{tabular}[c]{@{}c@{}}63.2 \\ (error $<$ sig. figs.)\end{tabular}  \\
Q$_{II}^{D}$ (MO)                                              & 3200                                                       & 950                                                         & 0.4                                                                  & \begin{tabular}[c]{@{}c@{}}103.6 \\ (error $<$ sig. figs.)\end{tabular}
\end{tabular}
\caption{The make up of simulations for the initial self-assembly of mesophases used in this study. Weight proportions calculated using the molecular masses, and for water are corrected to account for the Martini 4:1 mapping of water molecules \cite{Marrink2007}. Details of the calculation of parameters for the H$_{II}$ mesophase are in section S2 of the SI.}
\label{tab:SA_details}
\end{table}

\begin{figure}
    \centering
    \includegraphics[width=\linewidth]{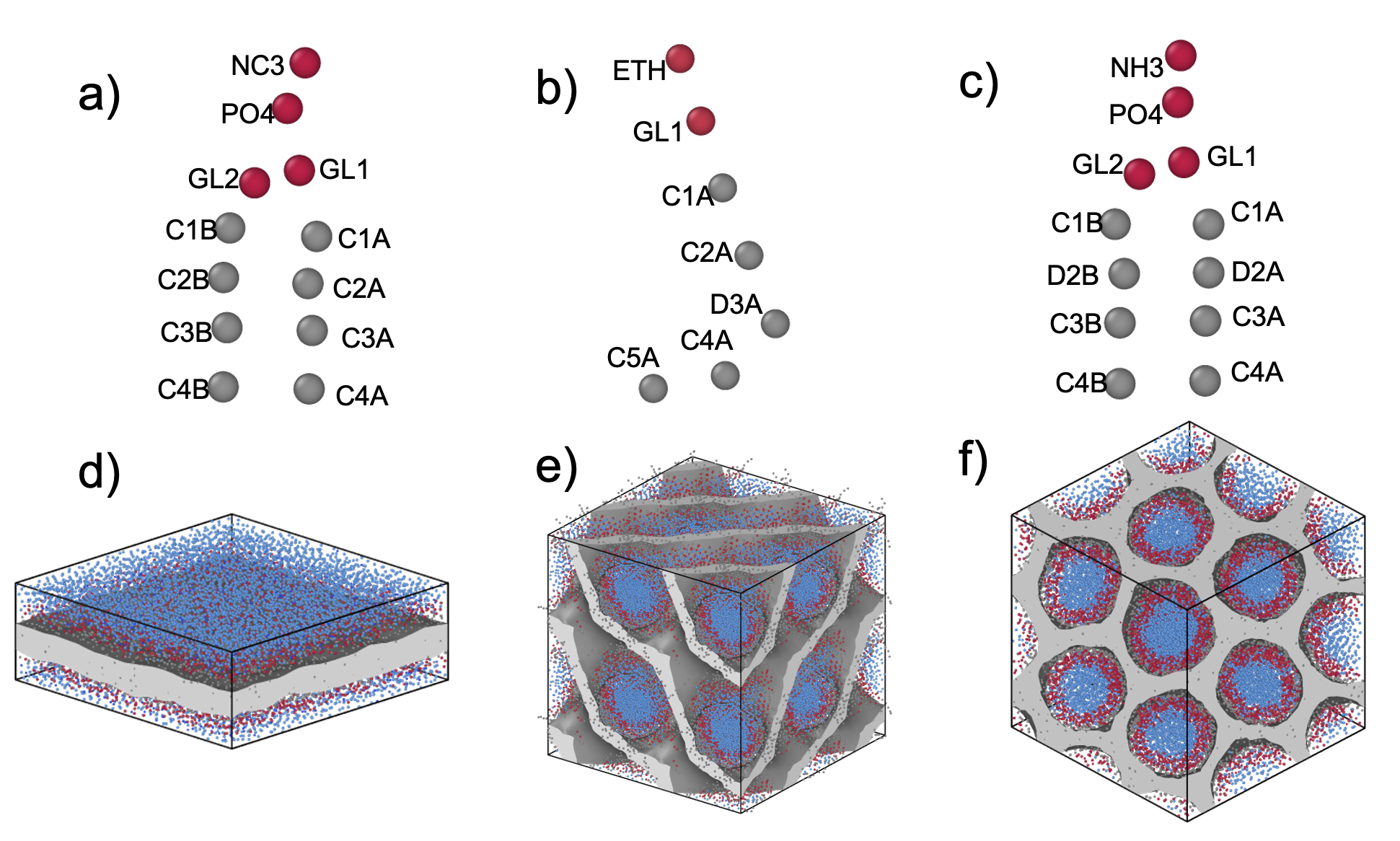}
    \caption{a)-c) coarse grained Martini models of DPPC, Monoolein, and DOPE respectively, with each atom labelled with its name used throughout the text. Headgroup atoms are Red, and carbons are Grey. d)-f) are simulation snapshots of the self-assembled mesophases, the L$_{\alpha}$, the Q$_{II}^{D}$, and H$_{II}$ respectively. In d)-f), water atoms are Blue, headgroup atoms Red, and carbon as in a)-c). For each mesophase, a grey surface has been constructed from the positions of the terminal carbon atoms using the method of \citeauthor{Stukowski2013} to show the mesophase more clearly \cite{Stukowski2013}.}
    \label{fig:molecules}
\end{figure}

The three mesophases in Figure \ref{fig:molecules} (Q$_{II}^{D}$, H$_{II}$, and L$_{\alpha}$) were simulated using the coarse-grained Martini force field version 2.1, using the standard simulation parameters for version 2, in Gromacs 2018 \cite{Marrink2007}. Coarse grained models of DOPE and DPPC were used from the standard Martini library \cite{Marrink2007}. Monoolein was modelled using the parameters of \citeauthor{Johner20141} \cite{Johner20141}.

The composition of each initial self-assembly simulation is detailed in table \ref{tab:SA_details}. The simulations were carried out at a pressure of 1 bar and temperature of 300 K. For the Q$_{II}^{D}$ mesophase, an initial equilibration simulation was necessary to ensure a well-mixed initial state before self-assembly\cite{Khelashvili2012, Fuhrmans2009}. 

The final frame of the short initial (150ns) self-assembly simulation was periodically replicated using Ovito. The initial H$_{II}$ and Q$_{II}^{D}$ unit cells were periodically replicated in the x, y, z directions 2 times for an expanded cell simulation of 2x2x2 unit cells. The initial L$_{\alpha}$ unit cell was expanded 3x3x1 times in the x,y,z directions for the final simulation cell. The expanded number of unit cells were then simulated for a further 1 $\mu$s (L$_{\alpha}$, H$_{II}$) or 1 ns (Q$_{II}^{D}$)\cite{ovito} for use to fit point clouds to. 11 (Q$_{II}^{D}$) or 34 (L$_{\alpha}$, H$_{II}$) equally spaced frames from the replicated cell simulations were used to measure final curvature distributions.

\subsection{Measuring curvature}

\begin{figure}
    \centering
    \includegraphics[width=.5\linewidth]{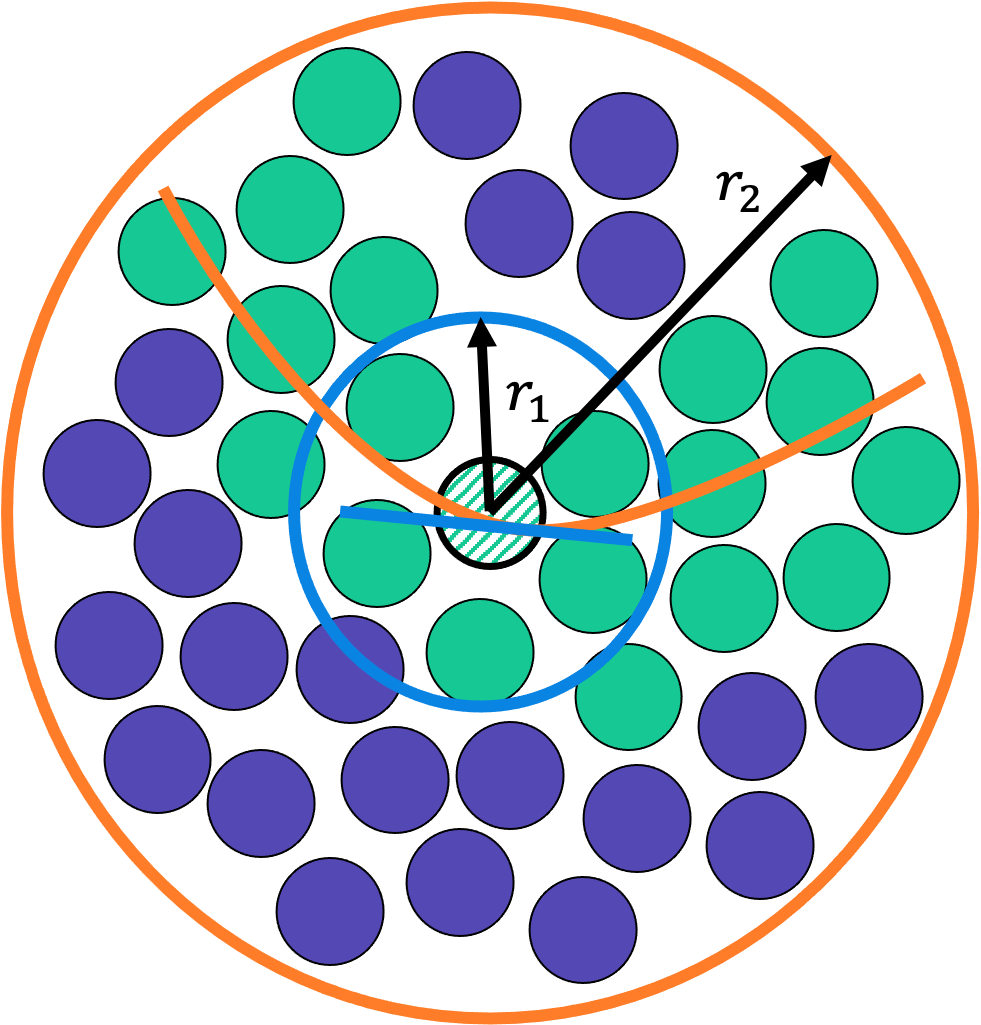}
    \caption{A 2 dimensional sketch of the method used, following \citeauthor{Yesylevskyy2014} \cite{Yesylevskyy2014}. A search is carried out on a selected green atom (shaded with a bold outline) for surrounding atoms of identical type at two different radii, $r_1$ (blue circle) and $r_2$ (orange circle), to create two different point clouds. After finding the principal axes of the point clouds, the atoms found are fitted to Equation \ref{eqn:surface}, the results of which are shown as two curves in the corresponding colours. Here, the green atoms could represent terminal molecule atoms, and the purple atoms (not used to form the point clouds), the atoms further towards the headgroups on either side of the bilayer.}
    \label{fig:method}
\end{figure}

The method of \citeauthor{Yesylevskyy2014} is illustrated in 2 dimensions in Figure \ref{fig:method}. Point clouds of atoms are generated by a particular cutoff radius from a selected atom in the system. Subsequently, the principal components of the cloud are calculated, and the entire cloud is transformed accordingly so that the smallest principal component is directed in the z direction. The cloud is then translated so that the selected atom is at $x=y=z=0$. The system is then fitted the generic quadric surface described by:

\begin{equation}
z = Px^2+Qy^2+Rxy+Sx+Ty+C
\label{eqn:surface}
\end{equation}

where $\{x,y,z\}$ are an orthogonal coordinate system and $\{P,Q,R,S,T,C\}$ are parameters to be optimised \cite{Yesylevskyy2014}. The fitted parameters are used to calculate the mean and Gaussian curvatures according to the fundamental forms of Equation \ref{eqn:surface}, as detailed in section S1 of the SI.

As the method described here uses point clouds of patches of membranes to measure curvature, it is possible to measure the curvature of a mesophase without prior knowledge or use of particular (non-Cartesian) coordinate system. This method will therefore allow the curvature properties of mesophase transitions to also be measured, and, as a local measure, has the potential to measure curvature preferences of molecules throughout simulations.

To begin this work, we used 5 frames from the initial (single cell) L$_{\alpha}$ simulation to compare the effectiveness of common fitting methods in the SciPy Python library to fit Equation \ref{eqn:surface} to point clouds. SciPy is a widely used scientific computing library in Python, which has a range of function optimisation routines\cite{2020SciPy-NMeth}. The fitting methods were \textit{Least Squares}, \textit{Minimize}, \textit{Dual Annealing}, \textit{Differential Evolution}, and \textit{Simplicial Homology Global Optimisation}.

\textit{Least Squares} is perhaps the most common and direct method used for finding the minima of many functions. In SciPy, the most common method is a Levenberg-Marquardt (LM) method, which uses a trust-region strategy for minimization. A trust-region method defines a region around the iterate in which the model can be trusted to be a good representation of the objective function. Using this region of the parameter search space, a step and direction are simultaneously chosen to continue to reduce the value of the objective function, until a tolerance is reached\cite{BFGS}.

\textit{Minimize} uses the BFGS Quasi-Newton method to find minima, by measuring changes in the gradient of the objective function at each iteration. Unlike LM, the BFGS method uses only an approximation of the Hessian matrix at each iteration step to determine the search direction, and so reducing the cost of calculation from $O(n^3)$ to $O(n^2)$, because no matrix-matrix operations are present in the method \cite{BFGS}.

\textit{Differential Evolution} (DE) and \textit{Dual Annealing} (DA) are stochastic optimisation methods, able to search a large parameter space for high-dimensional problems. Stochastic methods in this way trial potential solutions from a bounded initial parameter space, and develop further iterate solutions from candidate solutions throughout the iteration routine\cite{Xiang1997, Storn1997}.

\textit{Simplicial Homology Global Optimisation} (SHGO) is a recently-developed global optimisation technique, which, similarly to DE and DA works as a global optimisation algorithm, able to solve high-dimensional optimisation problems using input initial bounds on functions \cite{Endres2018}. However, unlike DE and DA, it does not use stochastic methods for finding minima in black-box input functions. Instead, SHGO finds multiple locally convex sub-domains of the objective function, which are individually fitted at each stage. Not only does this guarantee finding the global minimum of the function, if bounds are correctly chosen, but the location of local minima can also be simultaneously determined. This represents an improvement over stochastic methods, which may still converge to the same local minima if the initial sampling of the objective function is done poorly.

Once the fitting method had been established, curvatures were measured and averaged from frames of the expanded cell simulation, systematically varying the cutoff radius and measurement location. The reference Gaussian curvature of the cubic phase was calculated from the fitting of the implicit equation for the nodal approximation of the Diamond minimal surface, and using the parameters of the fitted surface after the method of \citeauthor{Goldman2005} \cite{PhysRevE.59.5528, Khelashvili2012, Goldman2005}.

The fitting code is available for open source use at \url{https://github.com/csbrasnett/lipid-md}

\section{Results}

\subsection{Effect of fitting method on curvature results}

\begin{figure}
    \centering
    \includegraphics[width = \linewidth]{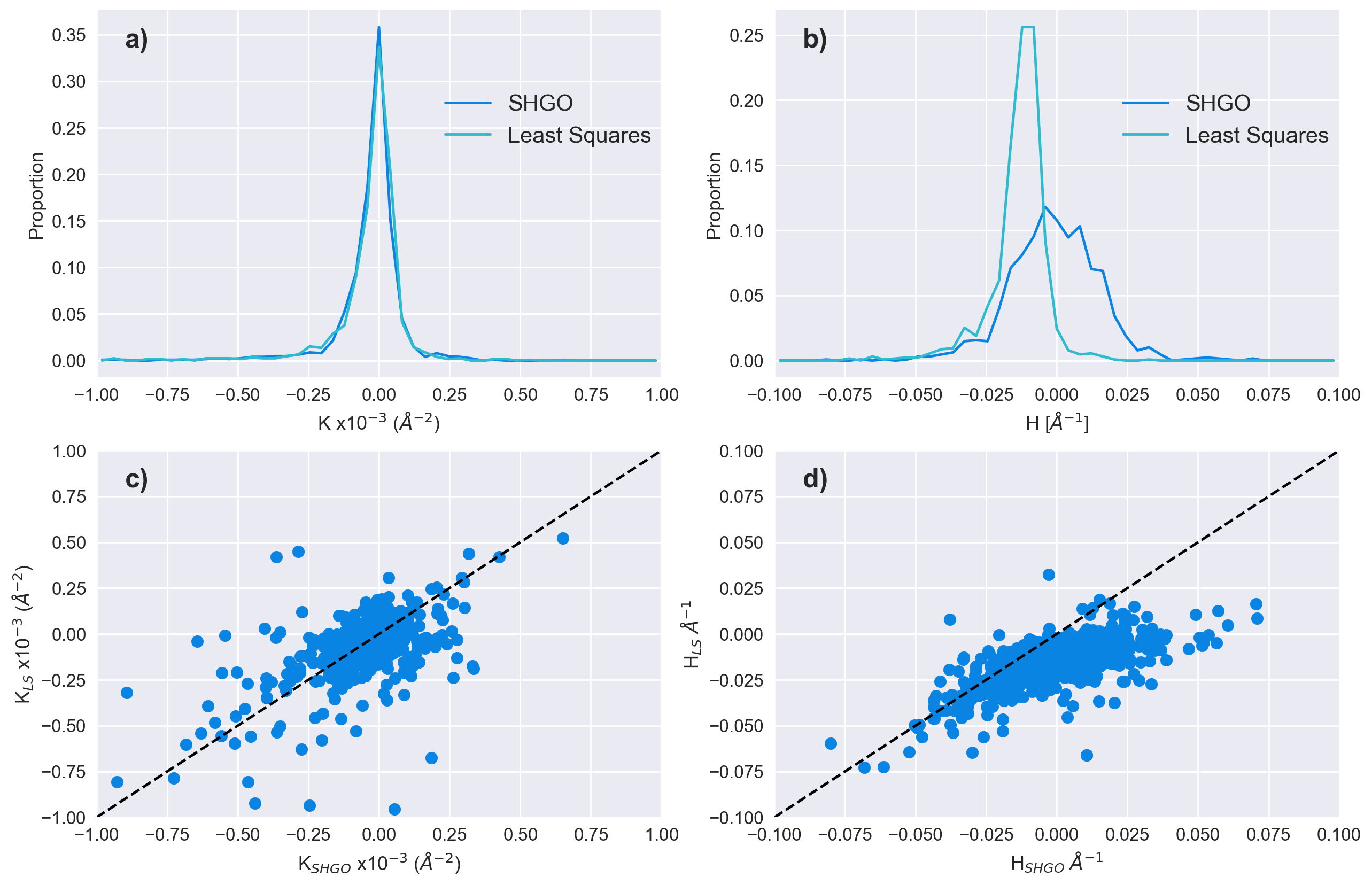}
    \caption{How the measures of Gaussian (a, c) and mean (b, d) curvature varies between the \textit{Simplicial Homology Global Optimization} and \textit{Least Squares} optimisation methods on an $L_{\alpha}$ mesophase at a cutoff radius of 25 \AA{}. a) the normalised distribution of Gaussian curvature between the two measures. The median values of the distributions are -5x$10^{-6}$ \AA{}$^{-2}$ and 1x$10^{-6}$ \AA{}$^{-2}$ for SHGO and \textit{Least Squares} respectively. b) the normalised distributions of mean curvature according to the different methods. The medians for SHGO and \textit{Least Squares} are 3x$10^{-18}$ \AA{}$^{-1}$ and -1x$10^{-2}$ \AA{}$^{-1}$ respectively. c) and d) show the correlation between the measurements of Gaussian and mean curvature respectively when fitting to identical point clouds, with black dashed lines indicating where perfect agreement between the methods lies. The distributions have R$^2$ values of 0.33 and 0.42 respectively. }
    \label{fig:LS_SHGO_comp}
\end{figure}

\begin{figure}
    \centering
    \includegraphics[width = \linewidth]{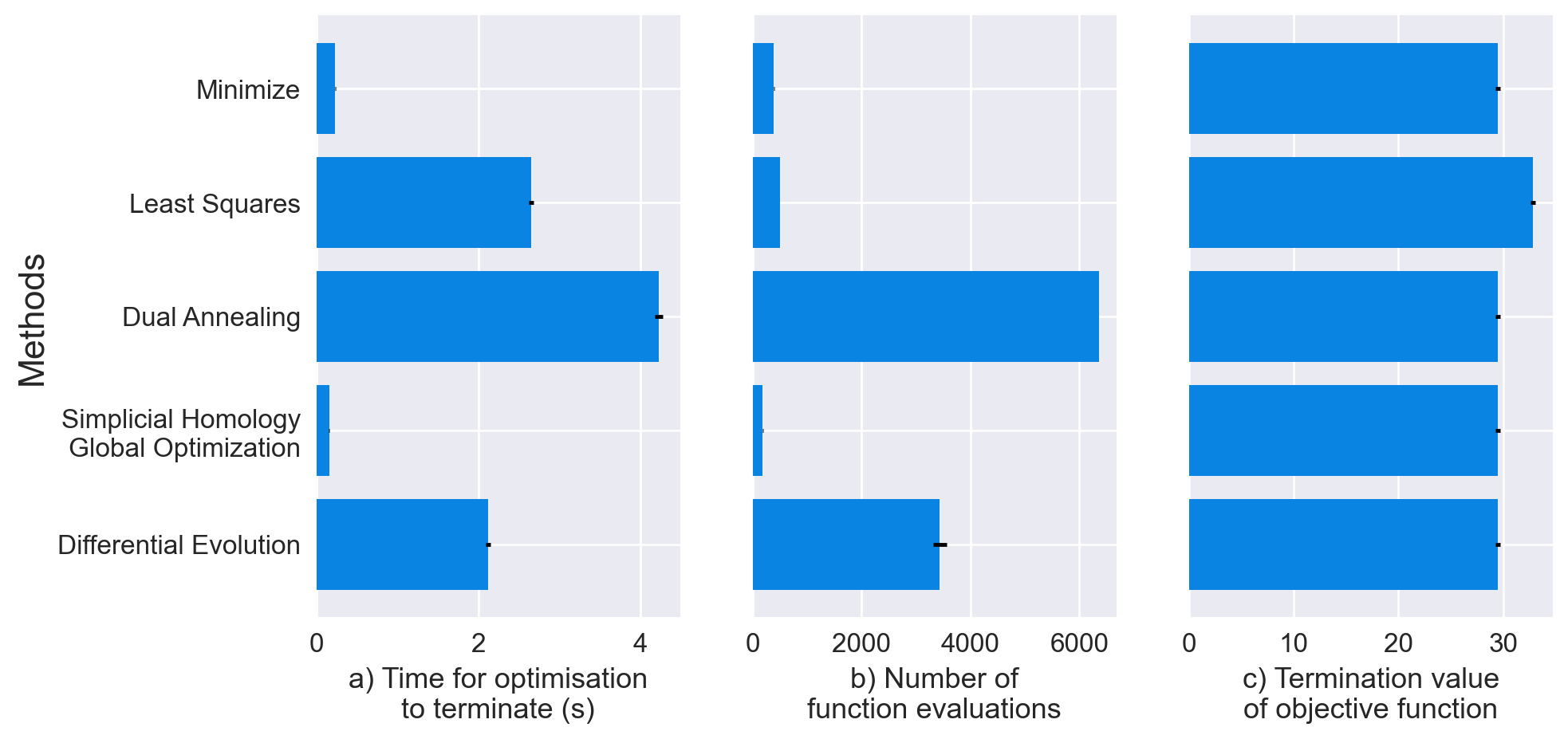}
    \caption{A comparison of the effectiveness of the fitting methods on a L$_{\alpha}$ mesophase, using a cutoff radius of 25 \AA{}. The data are evaluated across 1280 individual fits to point clouds across 5 simulation frames. The figures show a) the time for the optimisation routine to terminate, b) the total number of evaluations of the objective function that took place during the routine,  c) the value of the objective function upon termination}
    \label{fig:method_comparison}
\end{figure}

Our initial work compares the effectiveness of the 5 fitting methods described in the methods at fitting the surface described in Equation \ref{eqn:surface} to point clouds of atoms. For this, we used the terminal carbon beads (C4A and C4B as labelled in Figure \ref{fig:molecules}) of the $L_{\alpha}$ mesophase at a cutoff radius of 25 \AA{}. As the $L_{\alpha}$ mesophase is based on a flat plane, the results should show a distribution around 0 for both the mean and Gaussian curvature measures, by any measure of the curvature. Ideally, the surface fitted by each method would be identical, and the curvatures calculated would match exactly in every case. 

An example of the outcome of comparing these results in an $L_{\alpha}$ mesophase is seen in Figure \ref{fig:LS_SHGO_comp}, where we show the results for Gaussian and mean curvatures compared for the SHGO and \textit{Least Squares} methods close to 0. For a full comparison of all the fitting methods examined here, we have plotted every pair plot comparing methods in Figure S1 of the SI. In this region there is relatively low agreement between the results, as shown by the low R$^2$ values in Figure \ref{fig:LS_SHGO_comp} c) and d). However, there is still a remarkably close agreement between the overall distributions of the results for the Gaussian curvature, where both distributions are very narrowly distributed about 0. The agreement between the measures for mean curvature is slightly better, but the relatively sharp peak for \textit{Least Squares} centered at -1x$10^{-2}$ \AA{}$^{-1}$ shows that this measure can produce a strong systematic error for even a simple $L_{\alpha}$ mesophase. Taking Figure \ref{fig:LS_SHGO_comp} a) and b) together for \textit{Least Squares}, we can see that in general, the fit is producing cylindrical section surfaces, with K=0 and H$\neq$0. Comparing these two methods demonstrates the importance of using a wide trial parameter space and careful method selection in fitting point clouds of atoms. As we further show in Figures S1 and S2 of the SI, SHGO consistently produces the narrowest distributions of both mean and Gaussian curvature in comparison to any other method tried, and has closest agreement with the two other global fitting methods. In contrast, Least Squares fitting has a very poor level of agreement with any other method, and produces the results with the most extreme outliers.

In Figure \ref{fig:method_comparison}, we show that the choice of fitting method further effects the performance of the fitting routine. Figure \ref{fig:method_comparison} used 1280 fits to point clouds across a simulation of a single unit cell of an $L_{\alpha}$ mesophase, with the number of iterations of each method limited to 500. Our results then show the time that each method takes to complete within its termination limit, with SHGO taking the shortest, and \textit{Dual Annealing} taking the longest. However, more importantly, in b) and c) we compare the value of the objective function when the method terminates (regardless of successful termination), and the total number of evaluations of the objective function that took place in doing so. For an ideal fit, the value of the objective function should be 0, with the surface matching each points exactly. However, where there is agreement between 4 of the methods, the \textit{Least Squares} method has a consistently higher value, indicating a poorer fit. Of the other methods, SHGO matches the final value of the objective function, while also using the fewest evaluations of the function throughout the fit, and taking the quickest time. Furthermore, as we show in both Figure \ref{fig:LS_SHGO_comp} and S1 \& S2 of the SI, the median value of the mean and Gaussian curvatures in the L$_{\alpha}$ mesophase using SHGO is the closest to zero. Therefore, the final method for fitting point clouds from any mesophase uses the SHGO method, in order to ensure the most appropriate fit to the data.

\subsection{Effect of atom type on curvature measurement}

\begin{figure}
    \centering
    \includegraphics[width = \linewidth]{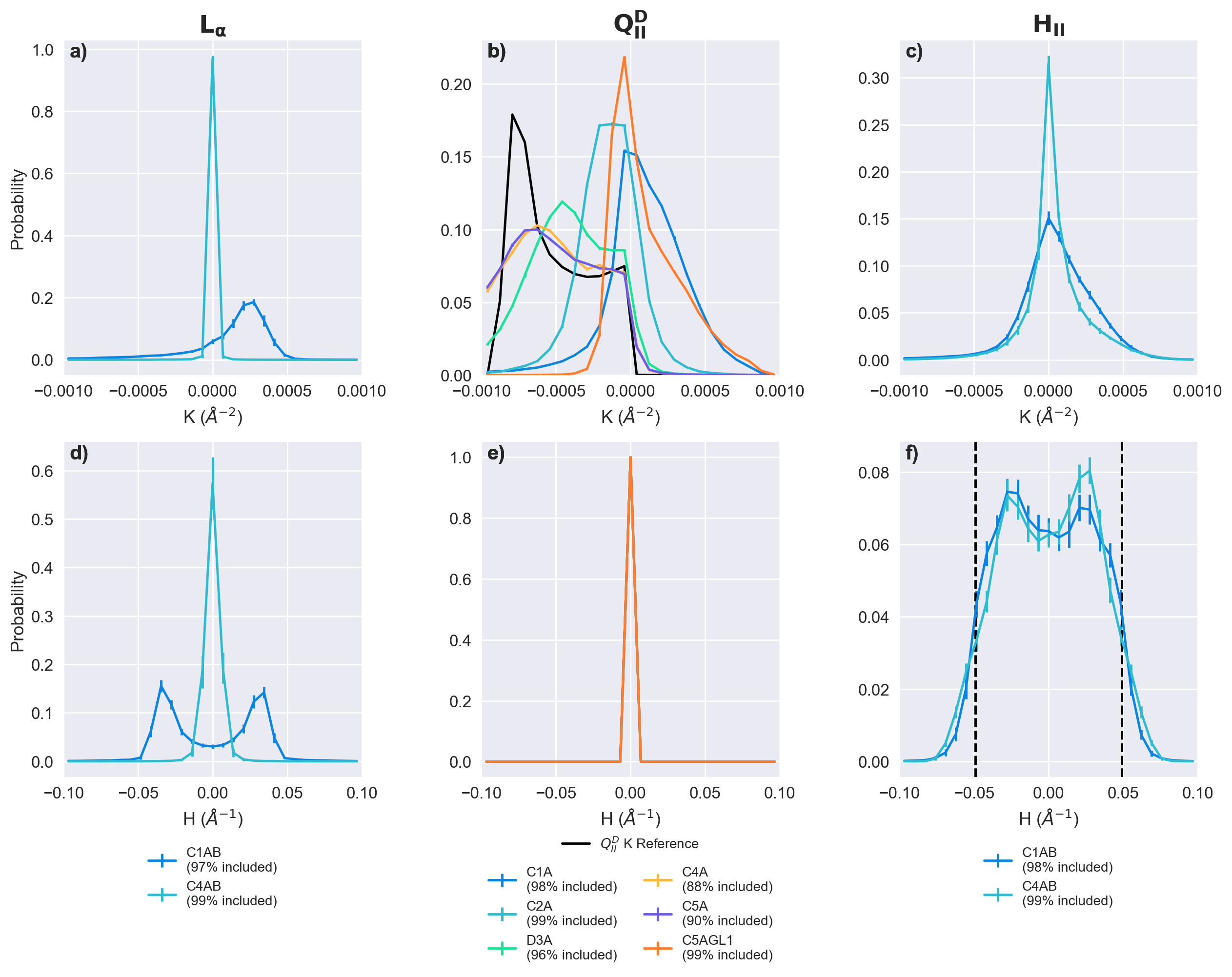}
    \caption{The distributions of (a-c) Gaussian and (d-f) mean curvatures in the $L_{\alpha}$ (a, d), $Q_{II}^{D}$ (b, e), and $H_{II}$ (c, f) mesophases, with the surface described in Equation \ref{eqn:surface} fitted at different locations in the membrane. For every fitting routine, the cutoff radius for point cloud formation was 40 \AA{}. Dotted lines in f) indicate the position of the expected peak in mean curvature based on measurements of the radius of the C1A cylinder in the simulation frames. Mean curvature is distributed about 0 as a feature of sign convention; negative mean curvature has no particular physical significance. The labels C1AB and C4AB indicate fits to point clouds formed of both the C1A and C1B atoms indicated in the molecule models in Figure \ref{fig:method}.}
    \label{fig:location_variation}
\end{figure}

Having established that SHGO was the optimal method for fitting Equation \ref{eqn:surface}, we investigated the effect of varying the location in the membrane at which to measure curvature, at a range of cutoff radii. For a radius of 40 \AA, we show results in Figure \ref{fig:location_variation}. Equivalent figures for other radii are in the SI.

For the L$_{\alpha}$ and H$_{II}$ trajectories, curvatures were measured at the initial (C1A, C1B) and terminal  (C4A, C4B) carbons as labelled in Figure \ref{fig:molecules}. For measurements of curvature in the Q$_{II}^{D}$ simulation, we measured curvature through independent point clouds of all 5 carbon atoms. Additionally, we trialled the method of \citeauthor{Yesylevskyy2014} in taking the positions of both the terminal carbons and headgroups, to fit a surface inbetween them.

In Figure \ref{fig:location_variation}, we firstly note the proportion of measured Gaussian curvature data which has been fitted in the range seen within the figure. This necessarily arises from comparison with the calibration for the Gaussian curvature in the Q$_{II}^{D}$ mesophase, where the distribution is known from the size of the simulated cell \cite{PhysRevE.59.5528}. While the proportion included in the measurements for the L$_{\alpha}$ and H$_{II}$ mesophases is close to 100\%, there is significant variation in the Q$_{II}^{D}$ measurements. The proportion fitted within the expected range is far greater when point clouds of atoms closer towards the centre of the bilayer are used. This data is itself plotted for cutoff radii in Figures S3-S5 of the SI, in which we also show that at shorter cutoff lengths, the proportion included in the expected range for the Q$_{II}^{D}$ mesophase measurements is significantly reduced.

In the L$_{\alpha}$ and H$_{II}$ mesophases, the Gaussian curvature is expected to be 0. The mean curvature of the L$_{\alpha}$ mesophase is also 0, but in the H$_{II}$ phase it is non-zero, as it is measured at the (cylindrical) pivotal plane. For an H$_{II}$ mesophase, the pivotal plane is located around the C1A/C1B atom location in the system, with a radius of 20.1 \AA{} for the H$_{II}$ system, used here\cite{Johner20140}. The results presented in Figure \ref{fig:location_variation} show that while Gaussian curvature is distributed around 0 as expected regardless of the location in the membrane it is measured, there is a more significant variation in the measurements of mean curvature. This is particularly evident in the case of the L$_{\alpha}$ mesophase, where taking the C1A/C1B beads of the system has resulted in a distribution of curvature not averaged about 0. However, for the H$_{II}$ mesophase, the measurement of mean curvature shows peaks close to the expected value of 0.049 \AA{}$^{-1}$ as indicated by the dashed black lines in Figure \ref{fig:location_variation} f). Further, at a cutoff of 30 \AA{} as we show in Figure S6, the distribution has peaks at the expected value.

For measurements of the L$_{\alpha}$ system, the discrepancy in mean curvature measurements is likely due to the chosen cutoff of 40 \AA{} in Figure \ref{fig:location_variation}. This cutoff radius is larger than the bilayer thickness, which results in both sides of the bilayer being captured to fit, and so a non-planar fit is usually found, for example in Figure S7 of the SI. Indeed, as seen in Figure S8 of the SI, at smaller cutoff radii of 10 \AA{}, the measurement of mean curvature in the L$_{\alpha}$ is averaged around 0, albeit with a comparatively broad distribution.

In comparison, the measurements of mean curvature in the Q$_{II}^{D}$ mesophase are distributed around 0 regardless of the location within the membrane to fit. The Q$_{II}^{D}$ surface is defined by the property of having zero mean curvature, so in this regard, the measurement is as expected. The measurement of Gaussian curvature in the Q$_{II}^{D}$ mesophase varies far more significantly. While no measurement conditions perfectly resemble the theoretical distribution of Gaussian curvature, using point clouds of terminal carbon atoms closely resembles it, with the vast majority of the data within the expected range for the unit cell size. As the Q$_{II}^{D}$ mesophase can be thought of as consisting of a lipid bilayer wrapped around an underlying minimal surface at the centre, this is expected.

Following the technique of \citeauthor{Yesylevskyy2014}, we additionally measured Gaussian curvature in the Q$_{II}^{D}$ phase using point clouds formed of both the terminal carbon C5A atom and the GL1 headgroup atom together. As Figure \ref{fig:location_variation} b) shows, this is perhaps the poorest measure of any. In this case, the first stage of the fitting routine often does not identify the principal axes of the point cloud correctly, and so Equation \ref{eqn:surface} cannot be fitted correctly to it (for example, see Figure S9 of the SI).

\subsection{Effect of cutoff radius on curvature measurement}

\begin{figure}
    \centering
    \includegraphics[width = \linewidth]{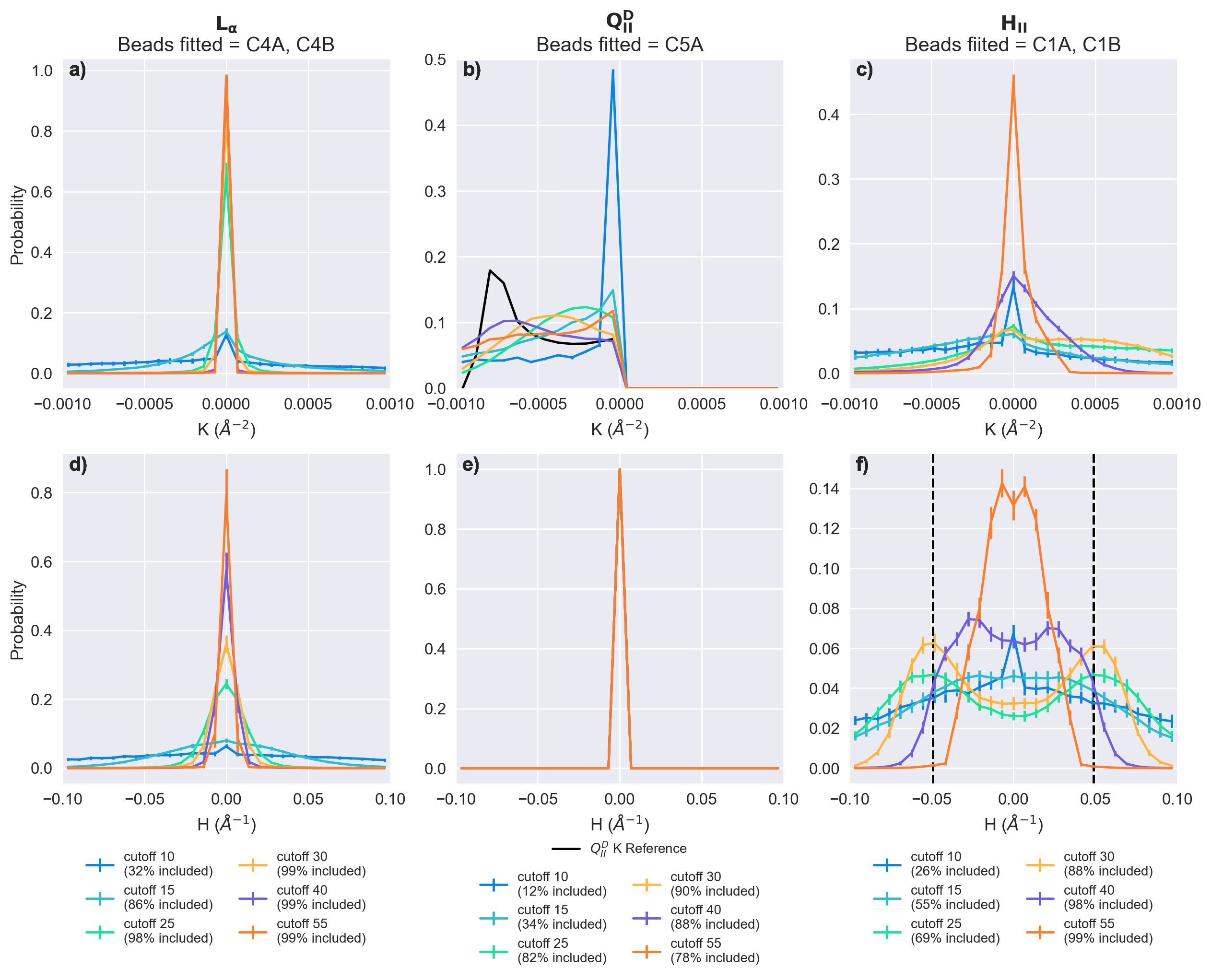}
    \caption{The (a-c) Gaussian and (d-f) mean curvature distributions at varying point cloud cutoff radii in the  L$_{\alpha}$ (a,d), H$_{II}$ (b,e), and Q$_{II}^{D}$ mesophases. In each case, Equation \ref{eqn:surface} has fitted to point clouds of the optimum fitting atoms as determined by the analysis in Figure \ref{fig:location_variation}. These were the C4A \& C4B, C5A, and C1A \& C1B atoms for the L$_{\alpha}$, H$_{II}$, and Q$_{II}^{D}$ mesophases respectively. Black dotted lines in f) indicate the fitted radius of the cylinder formed by the C1A and C1B atoms.}
    \label{fig:radius_variation}
\end{figure}

Given that curvatures vary widely with the location in the membrane where they are measured, we further systematically examined the effect of how the results vary with the cutoff radius used, which we show in Figure \ref{fig:radius_variation}. As might reasonably be expected, the results for the L$_{\alpha}$ mesophase show that as the cutoff radius increases, and the number of terminal carbons fitted to increases, both the mean and Gaussian curvatures become increasingly tightly distributed around the expected mean of 0. Although the same expected pattern is observed for the mean curvatures in the H$_{II}$ mesophase, the variation in the Gaussian curvature is far more significant. At both small and large cutoff radii, the measurements are distributed around 0, without the peaks at the appropriate values for the radius of the C1A, C1B cylinder indicated by the dashed lines. The fact that these peaks only emerge at intermediate cutoff radii indicates that at smaller and larger values, the surface being fitted becomes planar. At cutoff radii of 25 and 30 \AA{}, the peaks emerge at the expected value of H.

In comparison, the distribution of mean curvature in the Q$_{II}^{D}$ mesophase is very consistently a sharp peak around 0 irrespective of the cutoff radius. However more than any other, Figure \ref{fig:radius_variation} b) demonstrates that careful consideration needs to be given to the cutoff radius used to measure curvatures. While a cutoff of 40 \AA{} has a peak at more negative K as expected, at more extreme cutoffs, the peak completely disappears. As when varying the location in the membrane for the measurements, we also note that changing the cutoff radius has a significant effect in how much data fall within the expected range not just in the Q$_{II}^{D}$ mesophase, but also in the L$_{\alpha}$ and H$_{II}$ in Figure S10 of the SI.

\section{Conclusions}

In this work, we have presented a general method for analysing the mean and Gaussian curvatures of simulated lipid mesophases. We have evaluated how the method works under a variety of classic and recently developed optimisation methods, showing that optimisation methods designed for global optimisation perform significantly more self-consistently, and faster, than older methods that have been used previously. In particular, we have demonstrated the efficacy of the recently-developed SHGO method for a many-parameter problem such as the one used here \cite{Endres2018}. We subsequently showed that using the mesophase-agnostic method cannot be done naively: depending on the mesophase, the curvature distributions measured can vary significantly if improper choices are made for either the membrane location or the point cloud cutoff radius.

The validation of the mesophase independent method in this work will allow us in future to use molecular dynamics studies to inform experimental work. In particular, the study of lipid mesophase transitions is significant in the context of a number of biological processes, such as the formation of ceramide skin barriers, membrane fusion, and enzyme activity \cite{Wennberg2018, Narangifard2018, Siegel1990}. The method presented here will enable new understanding of the role that curvature plays in these transitions. Moreover, as a local measurement, the method will enable the tracking of individual molecules in mesophases, providing important contributions to questions around the curvature preference of different molecules.

\bibliography{ms}
\typeout{get arXiv to do 4 passes: Label(s) may have changed. Rerun}

\end{document}



\section{S1: Calculation of curvature}

For a surface defined by $z = f(x,y)$, we can simply parameterise the surface as $\vec{r}(u,v) = (u,v,f(u,v))$. The fitted surface in the main text is defined by the quadric form:

\begin{equation}
z = Px^2+Qy^2+Rxy+Sx+Ty+C
\label{eqn:surface}
\end{equation}

Using the parametric form of this surface, we can calculate the first ($I$) and second ($II$) fundamental forms of the surface, defined as:

\begin{equation}
    \begin{split}
        I = Edu^{2} + Fdudv + Gdv^{2}\\
        II = Ldu^{2} + 2Mdudv + Ndv^{2}
    \end{split}
\end{equation}

where:

\begin{align}
    E &= \vec{r}_u \cdot \vec{r}_u     & F &= \vec{r}_u \cdot \vec{r}_v & G &= \vec{r}_v \cdot \vec{r}_v\\
    L &= \vec{r}_{uu} \cdot \textbf{n}  & M &= \vec{r}_{uv} \cdot \textbf{n} & N &= \vec{r}_{vv} \cdot \textbf{n} \nonumber
\end{align}

where $\textbf{n}$ is the normal vector to the surface, itself defined by the initial parameterisation:

\begin{equation}
    \textbf{n} = \frac{\vec{r}_u \times \vec{r}_v }{\left | \vec{r}_u \times  \vec{r}_v \right |}
\end{equation}

Subsequently, noting that for the surfaces fitted to point clouds such that the point cloud-forming atom is at $x$=$y$=0, we arrive at the following values for the parameters:

\begin{align}
    E &= 1 + S^2    & F &= ST & G &= 1 + T^2\\
    L &= 2P   & M &= R & N &= 2Q \nonumber
\end{align}

Therefore, the metric tensors of the fundamental forms are:

\begin{equation}
I= \left( \begin{array}{ccc}
1 + S^2 & ST\\
ST & 1 + T^2\end{array} \right)\qquad
II= \left( \begin{array}{ccc}
2P & R\\ 
R & 2Q\end{array} \right)
\end{equation}

The mean (H) and Gaussian (K) curvatures are then defined by $I$ and $II$:

\begin{gather}
    K = \frac{det(II)}{det(I)}  = \frac{LN-M^2}{EG-F^2}\\
    H = \frac{1}{2}Trace((II)(I^{-1})) = \frac{LG-2MF+NE}{2(EG-F^2)} \nonumber
\end{gather}

\section{S2: Hexagonal mesophase calculations}

The lattice parameter. $a$ of an H$_{II}$ mesophase is defined by the distance between the centres of consecutive water channels. By simple geometry, we can therefore deduce:

\begin{equation}
    a = \sqrt{\frac{1}{\phi_w}(\frac{\pi}{2\sqrt{3}})d_w^2}
    \label{eqn:hex_a}
\end{equation}

where $\phi_w$ is the volume fraction of water in the system, and $d_w^2$ is the diameter of the water channels. $\phi_w$ is itself defined as:

\begin{equation}
    \phi_w = \frac{118N_w}{V_c}
\end{equation}

where $N_w$ is the number of water molecules in the system, $V_c$ is the volume of the simulation cell, and 118 \AA{} is the volume of 4 bulk water molecules \cite{Gerstein1996}.

The diameter of water channels was analysed using cluster analysis of particles in Ovito \cite{ovito}. Clusters of water atoms in single water channels were formed using cluster-forming cutoffs between 5.2 \AA{} and 10 \AA{}. The 5 largest clusters were individually fitted using a cylinder, and the radius of the fitted cylinder was used to determine the lattice parameter using Equation \ref{eqn:hex_a}. The radius was measured to be 14.2 $\pm$ 0.1 \AA{}, and $\phi_w$ was 0.29, so that the lattice parameter was 50.3 $\pm$ 0.3 \AA{}.

The same clustering construction was used to measure the radius of the ring formed by the C1A and C1B atoms in the mesophase to determine the expected value of the mean curvature at this location. Clusters were formed by using the water, NH3, PO4, GL1, GL2, C1A, C1B atoms, and cluster cutoff distances between 5 and 6 \AA{}. The radius was then measured as 20.1 $\pm$ 0.1 \AA{}, giving an expected value of H is 0.049 \AA{}$^{-1}$.

\section{S3: Effectiveness of fitting methods}

\begin{figure}
    \centering
    \includegraphics[height = 19cm]{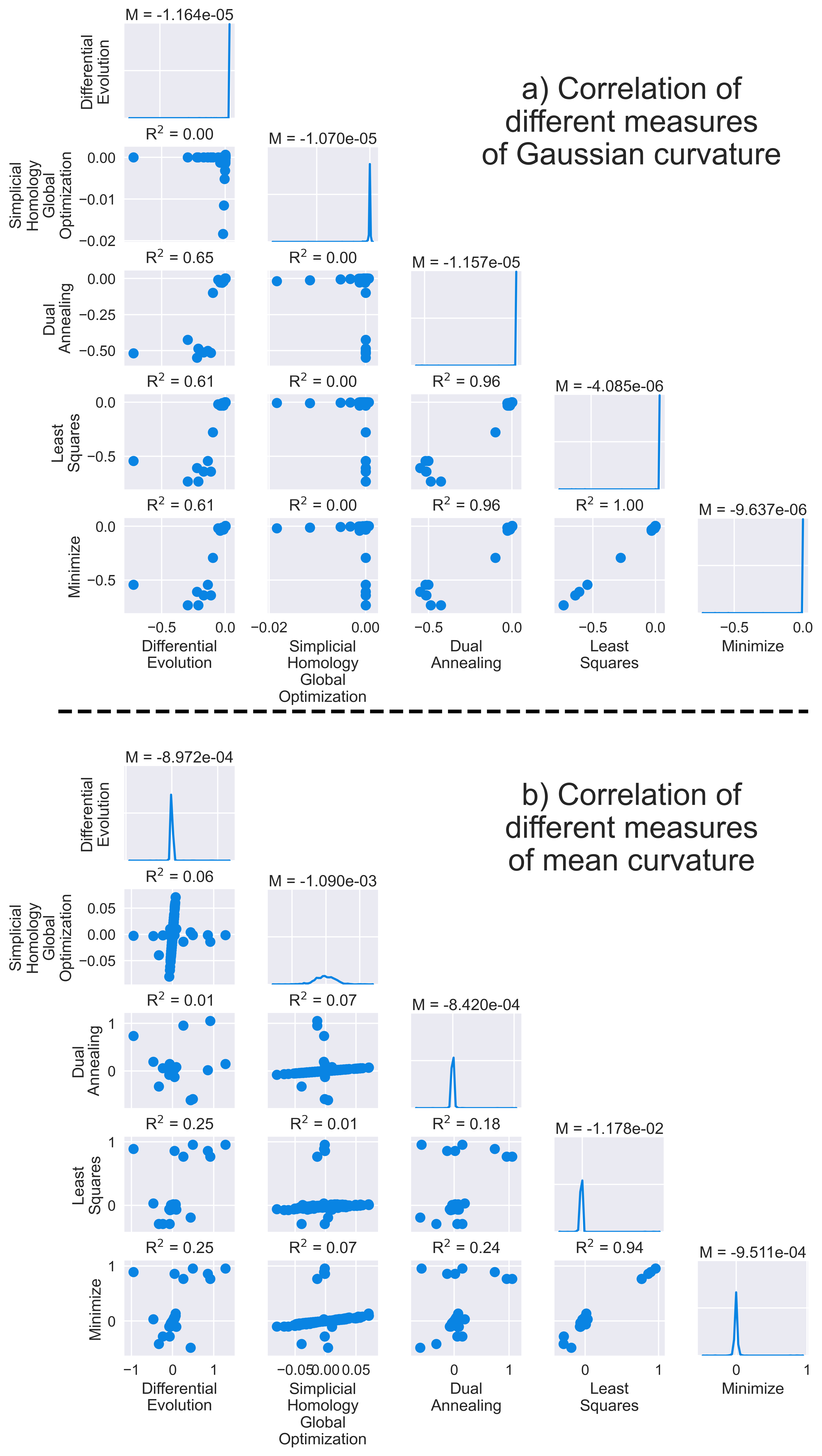}
    \caption{Pair plot comparisons of the values of mean and Gaussian curvature using 5 different fitting algorithms, the titles corresponding to the R$^2$ value of the data. The self-paired plots are histograms of the data, with the titles representing the median value. The lipid system the data have been fitted to is a L$_{\alpha}$ mesophase, and the cutoff radius for point cloud formation was 25 \AA{}.}
    \label{fig:fitter_comparison_1}
\end{figure}

\begin{figure}
    \centering
    \includegraphics[height = 20cm]{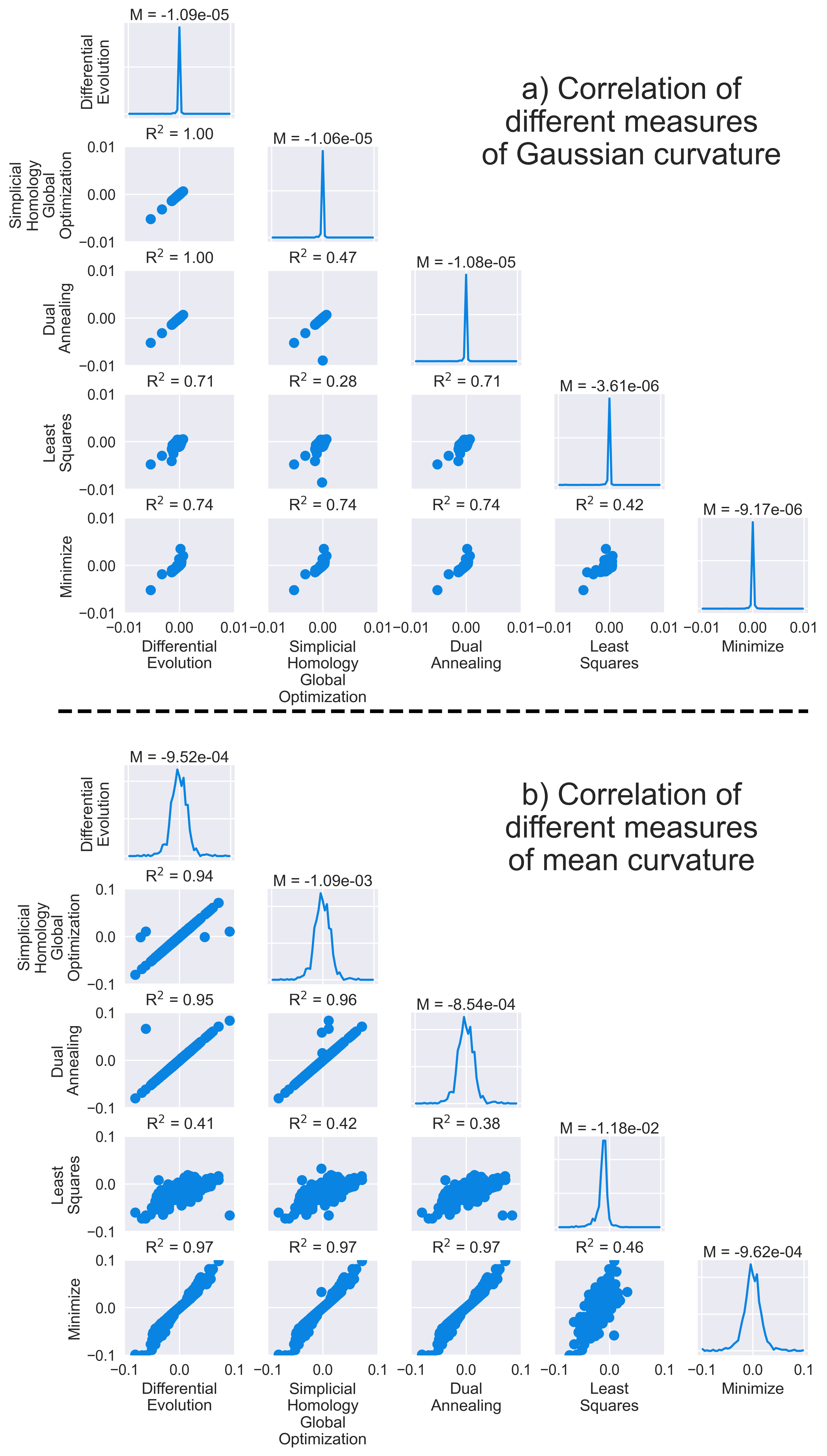}
    \caption{Highlights of the regions close to zero for the mean and Gaussian curvatures for the data in Figure \ref{fig:fitter_comparison_1}. Again, the R$^2$ and median values are in the titles of the plots, and there is a clear increase in comparison to the full dataset.}
    \label{fig:fitter_comparison_2}
\end{figure}

Figures \ref{fig:fitter_comparison_1} and \ref{fig:fitter_comparison_2} show full comparison plots between the fitting methods used for the quadric surface on a $L_{\alpha}$ mesophase at a cutoff of 25 \AA{}. Figure \ref{fig:fitter_comparison_2} shows a focus on the data are the regions of mean and Gaussian curvature close to 0, the expected result for a $L_{\alpha}$ mesophase.

\section{S4: Membrane location} 

\begin{figure}
    \centering
    \includegraphics[width = \linewidth]{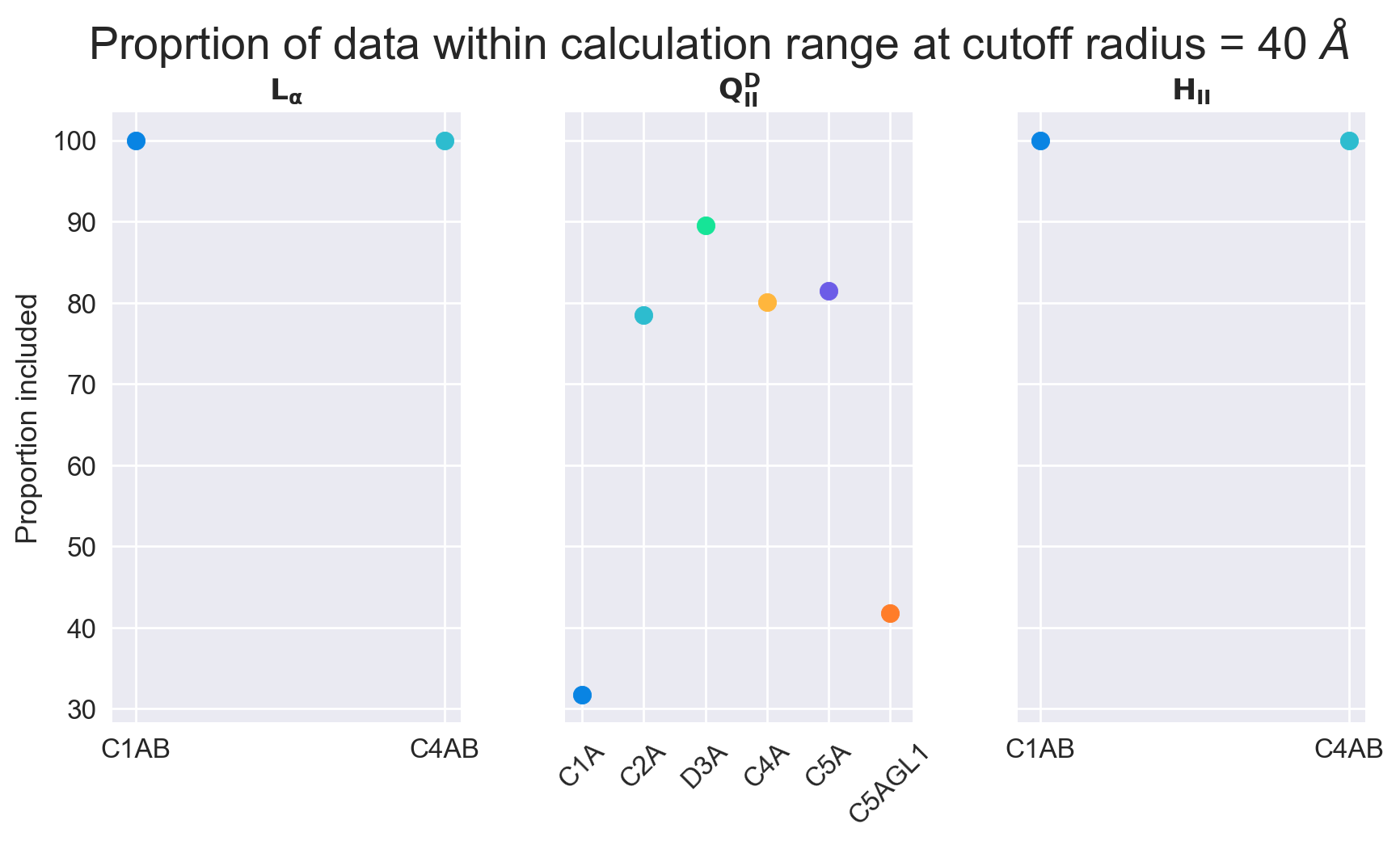}
    \caption{The proportion of data included within the expected range for the three mesophases proportion of data included within the expected range for the three mesophases measured, at different locations in the membrane, and at a fixed cutoff radius of 40 \AA{}. For the $L_{\alpha}$ and H$_{II}$ systems, the labels 'C1AB' and 'C4AB' indicate measures taken where the C1A and C1B, and C4A and C4B atoms respectively have been combined.}
    \label{fig:data_inclusion_40}
\end{figure}

\begin{figure}
    \centering
    \includegraphics[width = \linewidth]{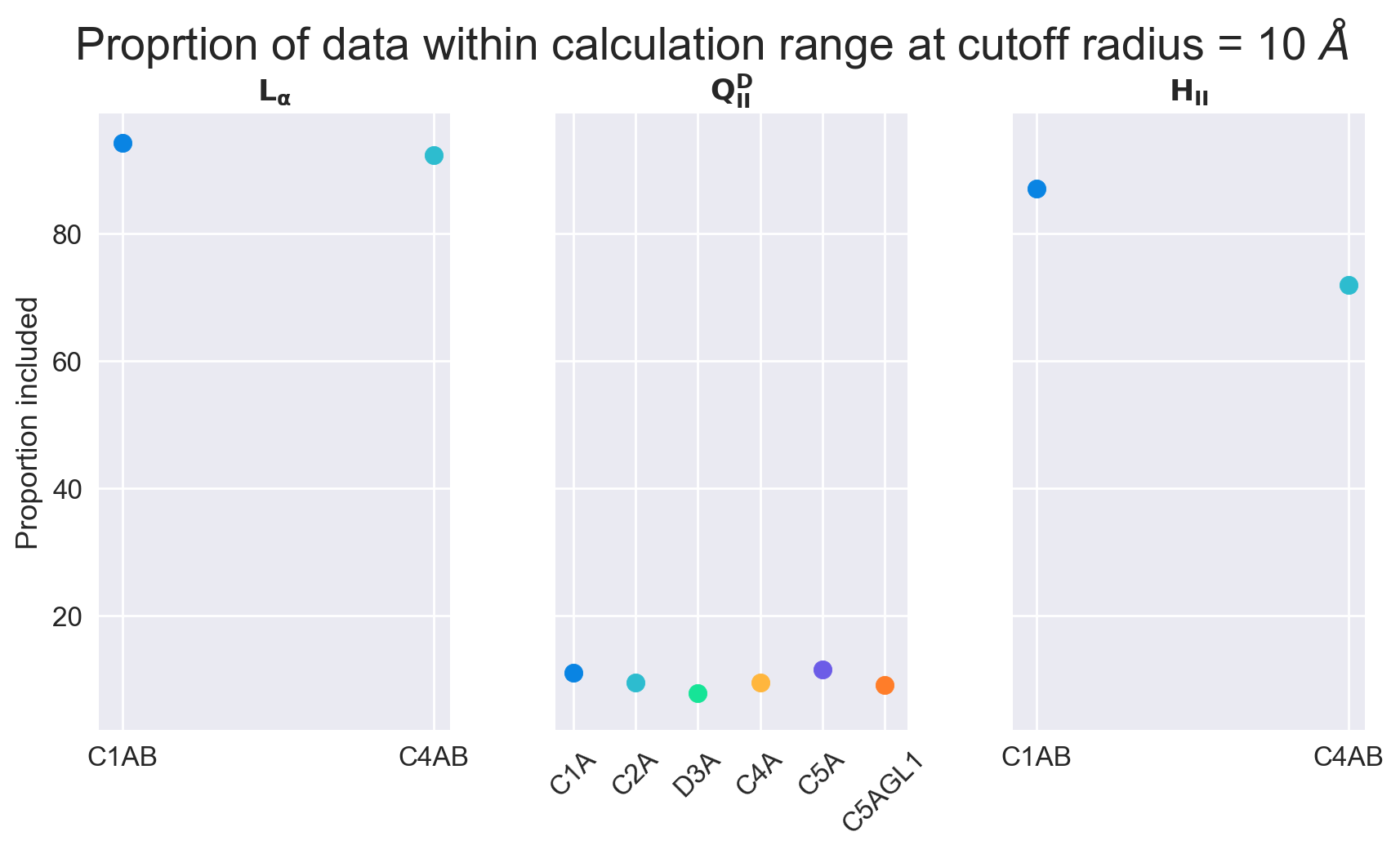}
    \caption{As with Figure \ref{fig:data_inclusion_40}, the variation of the proportion of data within the expected range at different locations within the membrane, at a cutoff radius of 10 \AA{}.}
    \label{fig:data_inclusion_10}
\end{figure}

\begin{figure}
    \centering
    \includegraphics[width = \linewidth]{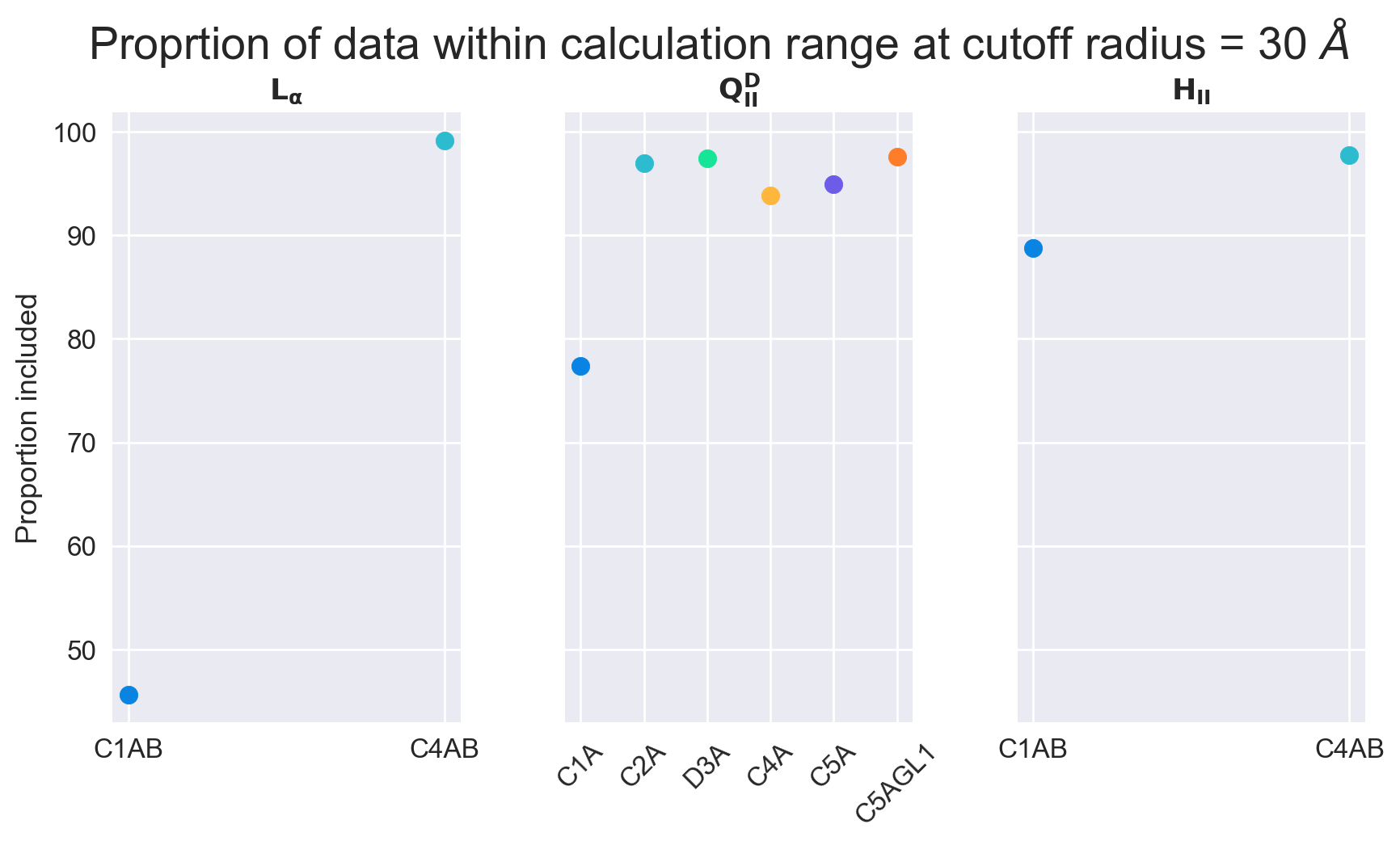}
    \caption{As with Figure \ref{fig:data_inclusion_40}, the variation of the proportion of data within the expected range at different locations within the membrane, at a cutoff radius of 10 \AA{}.}
    \label{fig:data_inclusion_30}
\end{figure}

\begin{figure}
    \centering
    \includegraphics[width = \linewidth]{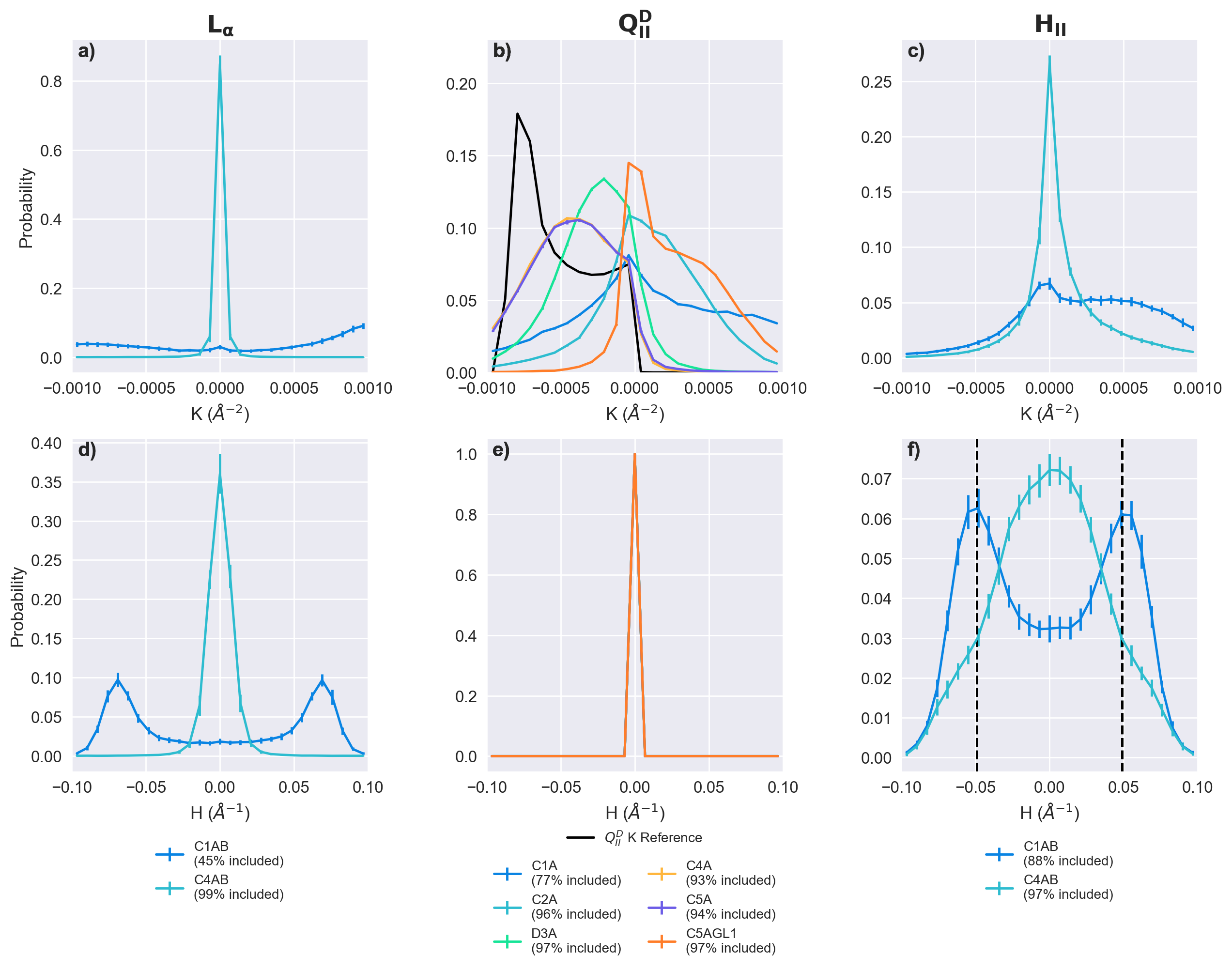}
    \caption{The distribution of Gaussian (a-c) and mean (d-f) curvatures as measured in the three trial mesophases at a cutoff radius of 30 \AA{}, at different locations within the membrane. }
    \label{fig:location_variation_30}
\end{figure}

\begin{figure}
    \centering
    \includegraphics[width = \linewidth]{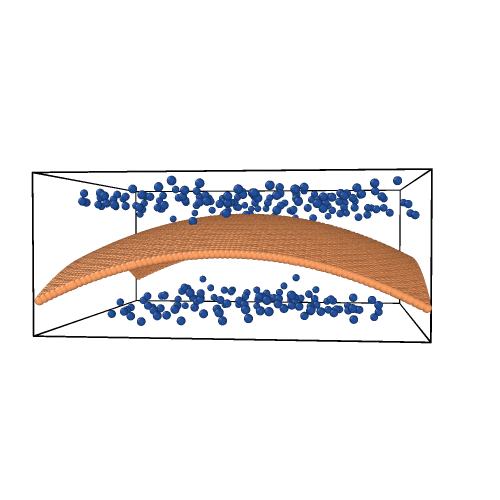}
    \caption{A point cloud of C1A and C1B atoms (blue), and a fitted surface (orange) in an $L_{\alpha}$ mesophase, at a cutoff of 40 \AA{}. The fitted surface clearly has positive (ie. non-zero) mean and Gaussian curvatures.}
    \label{fig:bad_La}
\end{figure}

\begin{figure}
    \centering
    \includegraphics[width = \linewidth]{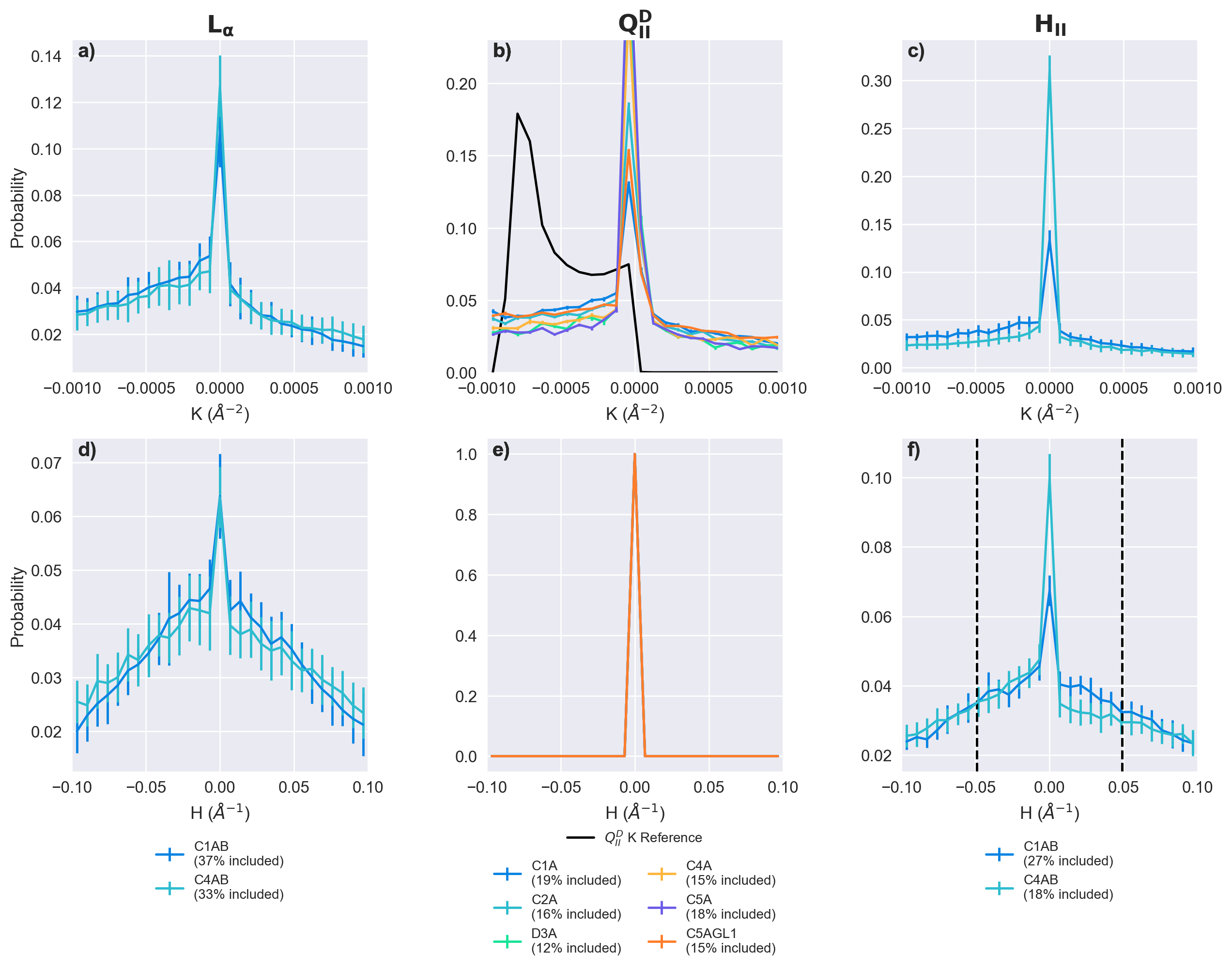}
    \caption{The distribution of Gaussian (a-c) and mean (d-f) curvatures as measured in the three trial mesophases at a cutoff radius of 10 \AA{}, at different locations within the membrane. }
    \label{fig:location_variation_10}
\end{figure}

\begin{figure}
    \centering
    \includegraphics[width=\linewidth]{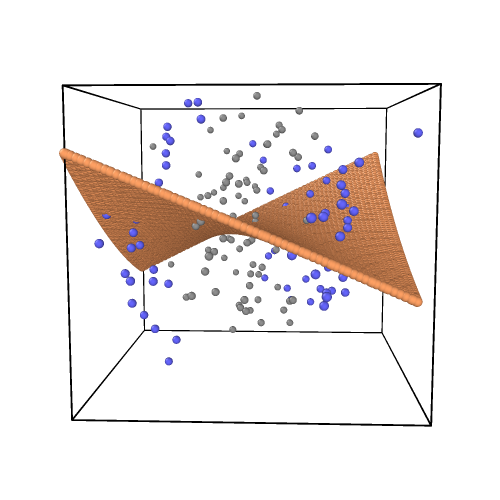}
    \caption{An attempt at fitting a point cloud comprised of C5A (grey) and GL1 (blue) atoms in an MO Q$_{II}^{D}$ mesophase. Performing an initial PCA on the cloud has not identified the surface normals correctly, and so the fitted surface is not representative of the membrane curvature at this point.}
    \label{fig:bad_C5AGL1}
\end{figure}

Figures \ref{fig:data_inclusion_40} and \ref{fig:data_inclusion_10} show that there is a significant effect on the proportion of data that lies within the expected range, depending on the location in the membrane used. The location within the membrane is indicated by the atoms fitted to. Additionally, there is a significant difference in the proportion included depending on the cutoff radius used to form the point cloud of atoms, as demonstrated by the difference between the two figures. Figure \ref{fig:bad_La} hints at the cause: in an $L_{\alpha}$ mesophase, forming point clouds of C1A and C1B atoms at a cutoff radius of 40 \AA{} begins to include atoms from both sides of the bilayer, so that the fitted surface has unexpected curvature. 
The full results for a cutoff radius of 10 \AA{} are shown in Figure \ref{fig:location_variation_10}. Compared to the equivalent figure for cutoff = 40 \AA{} in the main text (Figure 3), there is a significant broadening of the distributions of mean curvature, and in the cases of the $L_{\alpha}$ and H$_{II}$ systems, the Gaussian curvature as well. The Gaussian curvature for Q$_{II}^{D}$ compares extremely poorly, with little resemblance to the reference distribution. As Figure \ref{fig:data_inclusion_10} corroborates, there is also exceptionally little data within the expected measurement range.

\section{S5: Radius variation}

\begin{figure}
    \centering
    \includegraphics[width = \linewidth]{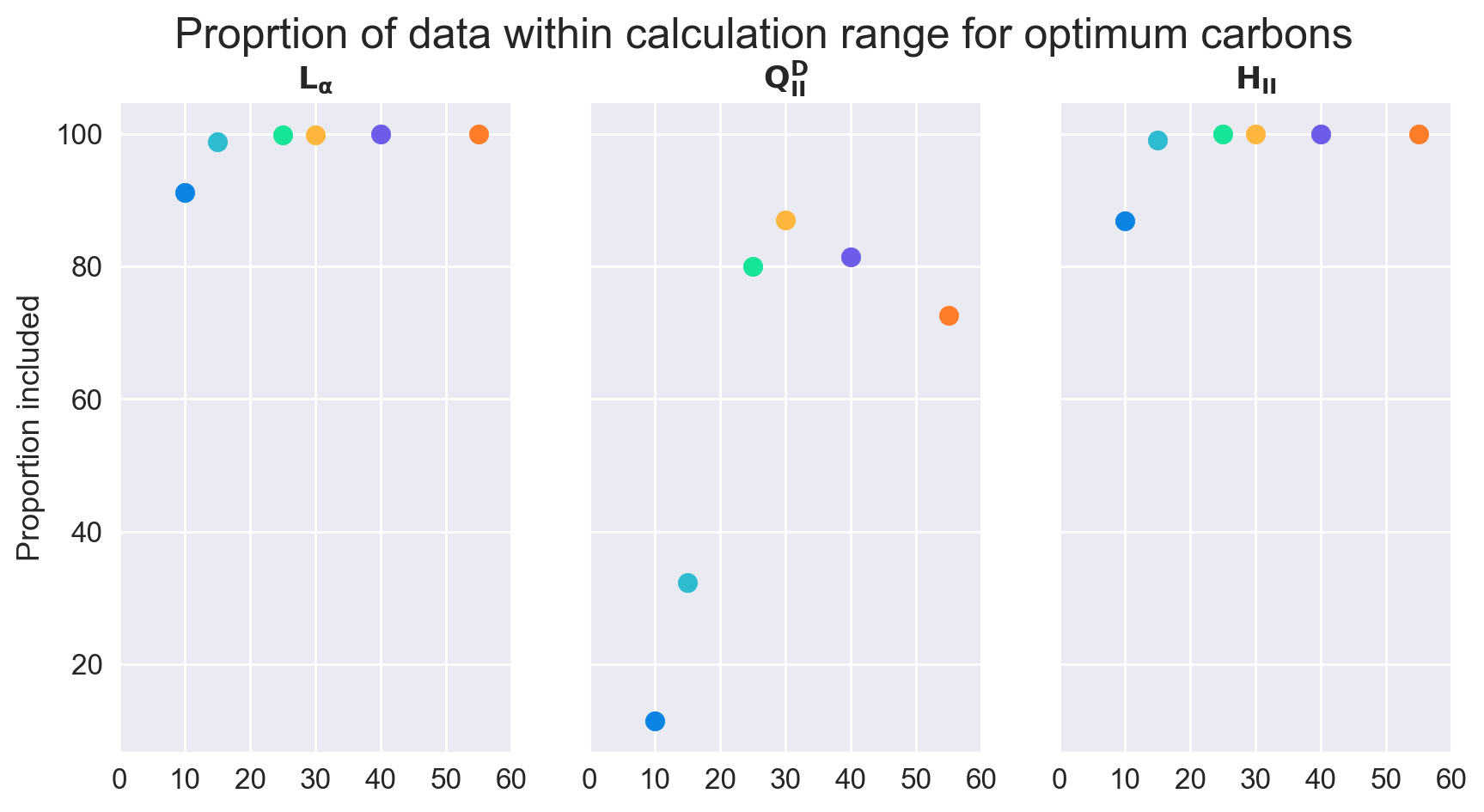}
    \caption{The proportion of data included within the expected range when the cutoff radius is varied at the optimum fitting location as determined by section S3.}
    \label{fig:radius_inclusion_terminalC}
\end{figure}

After accounting for the variation in fit due to the location in the membrane, Figure \ref{fig:radius_inclusion_terminalC} shows that there is again significant variation in the proportion of data fitted to within the expected range due to the cutoff radius. In the L$_{\alpha}$ and H$_{II}$ mesophases the effect is only really observed at shorter radii, while in the Q$_{II}^{D}$ mesophase, there is a clear optimum peak at 30 \AA{}.

\bibliography{supplement}
\typeout{get arXiv to do 4 passes: Label(s) may have changed. Rerun}